\documentclass{article}

\usepackage{arxiv}
\usepackage[utf8]{inputenc} 
\usepackage[T1]{fontenc}    
\usepackage{booktabs}       
\usepackage{amsfonts}       
\usepackage{nicefrac}       
\usepackage{microtype}      
\usepackage{lipsum}		
\usepackage{amsmath,amssymb}
\usepackage{graphicx}
\usepackage{color}

\newcommand{\grl}{    {Geophys. Res. Lett.}}
\newcommand{\jgr}{    {J. Geophys. Res.}}
\newcommand{\ssr}{    {Space Sci. Rev.}}
\newcommand{\planss}{    {Plan. Sp. Sci.}}

\newcommand{\solphys}{ {Solar Physics}}
\newcommand{\apj}{ {Astrophys. J. }}
\newcommand{\apjl}{    {Astrophys. J. Lett.}}

\newcommand{\mnras}{ {Mon.Not.Royal.Soc. }}

\newcommand{\prl}{    {Phys. Rev. Lett.}}
\newcommand{\apjs}{    {Astrophys. Journal. Suppl. Ser.}}

\newcommand{\blue}{\textcolor{black}}

\def\XXint#1#2#3{{\setbox0=\hbox{$#1{#2#3}{\int}$}
     \vcenter{\hbox{$#2#3$}}\kern-.5\wd0}}

\title{Kinetic-scale current sheets in the solar wind at 1 AU: Scale-dependent properties and critical current density}

\author{
    {Ivan Y. Vasko} \\
	Space Sciences Laboratory, University of California at Berkeley, California, CA94720, USA\\
	Space Research Institute of Russian Academy of Sciences, Moscow, 117997, Russia\\
	\texttt{vaskoiy@berkeley.edu} \\
	\and
	{Kazbek Alimov} \\
	Space Sciences Laboratory, University of California at Berkeley, California, CA94720, USA\\
	\and
	{Tai Phan} \\
	Space Sciences Laboratory, University of California at Berkeley, California, CA94720, USA\\
	\and
	{Stuart D. Bale} \\
	Space Sciences Laboratory, University of California at Berkeley, California, CA94720, USA\\
	Department of Physics, University of California at Berkeley, California, CA94720, USA\\
	\and
	{Forrest S. Mozer} \\
	Space Sciences Laboratory, University of California at Berkeley, California, CA94720, USA\\
    \and
	{Anton V. Artemyev} \\
	Institute of Geophysics and Planetary Sciences, University of California, Los Angeles, California, CA 90095, USA\\
	Space Research Institute of Russian Academy of Sciences, Moscow, 117997, Russia
}

\begin{document}
\maketitle

\begin{abstract}
We present analysis of 17,043 \blue{proton kinetic-scale} current sheets collected over 124 days of Wind spacecraft measurements \blue{in the solar wind} at 11 Samples/s magnetic field resolution. The current sheets have thickness $\lambda$ from a few tens to one thousand kilometers with typical value around 100 km \blue{or from about 0.1 to 10$\lambda_{p}$ in terms of local proton inertial length $\lambda_{p}$.} We found that the current density is larger for smaller scale current sheets, $J_0\approx 6\; {\rm nA/m^2} \cdot (\lambda/100\;{\rm km})^{-0.56}$ , but does not statistically exceed critical value $J_A$ corresponding to the drift between ions and electrons of local Alv\'{e}n speed. The observed trend holds in normalized units, $J_0/J_{A}\approx 0.17\cdot (\lambda/\lambda_{p})^{-0.51}$. The current sheets are statistically force-free with magnetic shear angle correlated with current sheet spatial scale, $\Delta \theta\approx 19^{\circ}\cdot (\lambda/\lambda_{p})^{0.5}$. The observed correlations are consistent with local turbulence being the source of proton kinetic-scale current sheets in the solar wind, while mechanisms limiting the current density remain to be understood.
\end{abstract}


\section{Introduction}


The understanding of turbulence dissipation
and plasma heating in a weakly collisionless plasma is
of fundamental importance for numerous astrophysical
systems \cite{Matthaeus&Velli11}. Numerical simulations showed that turbulence dissipation should be spatially intermittent with substantial plasma heating localized in and around coherent structures, such as current sheets, which occupy a relatively small volume \cite{Karimabadi13:phpl,Zhdankin13:apjl,Zhdanin14:apj,Wan14:apj,Wan16:phpl}. \blue{Numerical simulations also showed that magnetic reconnection in kinetic-scale current sheets produced by turbulence not only results in plasma heating, but also fundamentally affects development of the turbulence cascade at sub-proton scales \cite{Servidio10:phpl,Franci17:apj,Cerri&Califano17,Papini19:apj}}. Modern simulations are still incapable of entirely reproducing the complex dynamics of realistic three-dimensional plasma turbulence, and substantial effort has been directed toward comparing simulation results with observations in the solar wind, a weakly collisional plasma most accessible for {\it in-situ} measurements \cite{Matteini20:frph}. 

\blue{Spacecraft measurements showed that solar wind heating should continuously occur within a few tens of solar radii of the Sun as well as further out in the heliosphere and dissipation of turbulent magnetic field fluctuations should be the dominant solar wind heating mechanism \cite{Kohl96:apj,Cranmer09,Hellinger13:jgr}.} The solar wind observations revealed turbulence to be dominated by Alf\'{e}nic fluctuations highly oblique ($k_{\perp}\gg k_{||}$) to local mean magnetic field, Kolmogorov-like spectrum $E_{k_{\perp}}\propto k_{\perp}^{-5/3}$ at scales larger than proton kinetic scales, and a steeper spectrum $E_{k_{\perp}}\propto k_{\perp}^{-2.8}$ at scales smaller than proton kinetic scales \cite{Chen16:jpp}. Numerical simulations have successfully reproduced these properties of the solar wind \blue{turbulence \cite{Boldyrev&Perez12,Cerri17:apjl,Franci17:apj,Franci18:apj,Papini19:apj}}. The crucial unresolved problem is the origin of coherent structures and, specifically, current sheets observed in the solar wind, and the very ability of these structures to heat solar wind plasma \cite{Matthaeus15}. In this Letter we present a statistical analysis of proton kinetic-scale current sheets (CS) in the solar wind at 1 AU, contributing to understanding the origin and dissipation of coherent structures in plasma turbulence.

The presence of CSs on a wide range of temporal scales was established by early spacecraft measurements in the solar wind \cite{burlaga77,tsurutani79,lepping86,soding01}. These measurements showed that the magnetic field has more or less constant magnitude, but rotates across a CS through some shear angle. The typical CS thickness was around ten proton inertial lengths, while the occurrence rate at 1 AU was a few tens per day. The multi-spacecraft analysis showed that the magnetic field component normal to a CS surface is much smaller (if present at all) than local magnetic field magnitude \cite{Horbury01,Knetter04}. The majority of CS studies were limited by magnetic field measurements of relatively low resolution (a few seconds at best) and typically by CSs with shear angles $\gtrsim 30^{\circ}$ \cite{soding01,Artemyev18:apj,Artemyev19:grl}. Devoid of these shortcomings, the analysis by \cite{Vasquez07} included more than 6,000 CSs collected at 1 AU using magnetic field measurements at 1/3 s resolution. The typical thickness of the CSs was around a few proton inertial lengths, and the occurrence rate was about a few hundred CSs per day. Thus, the higher temporal resolution allowed resolving thinner CSs, which turned out being much more abundant than larger-scale CSs reported in the early studies. Based on the log-normal distribution of waiting times of the CSs, \cite{Vasquez07} suggested they are produced by local turbulence, in accordance with earlier theoretical hypothesis \cite{Matthaeus&Lamkin86}.  

This hypothesis was further supported by observations of similar distributions of magnetic field rotations and waiting times of coherent structures in the solar wind and MHD turbulence simulations \cite{Greco08,Greco09:apjl,Zhdankin12:apjl}. One of the alternatives is that magnetic field rotations through angles $\gtrsim 30^{\circ}$ are boundaries between flux tubes originating at the Sun, while smaller rotations are produced by local turbulence \cite{Bruno01:pss,Borovsky08:jgr}. However, this interpretation can hardly explain the universal log-normal distribution of magnetic field rotations at various temporal scales \cite{Zhdankin12:apjl,Chen15:mnras}. The ability of coherent structures in heating of solar wind plasma was questioned by \cite{borovsky11}, but there is currently a growing evidence that plasma heating does occur around coherent structures in plasma turbulence \cite{Osman11:apjl,Osman12b,Chasapis15:apjl,Chasapis18}. 

In this Letter we present the most extensive dataset of proton kinetic-scale CSs collected at 1 AU using magnetic field measurements at 1/11 s resolution. \blue{Due to the higher resolution, the CSs with thickness from a few to ten times smaller than in the previous studies could be resolved.} We reveal distinct scale-dependencies of the current density and shear angle, as well as the critical current density \blue{that is not exceeded statistically}. The results indicate that \blue{proton kinetic-scale CSs} in the solar wind are indeed produced by \blue{turbulence cascade}, and advance understanding of turbulence dissipation.




\section{Data and methodology \label{sec:2}}
 
We use measurements of the Wind spacecraft that is located at the L1 Lagrangian point, about $200$ Earth radii from the Earth \cite{Wilson21}. We use continuous magnetic field measurements at $1/11$ s resolution provided by the Magnetic Field Instrument \cite{Lepping95:ssr}, and proton densities and flow velocities at 3s cadence provided by Wind 3DP instrument \cite{Lin95}. Note that ion densities coincide within a few tens of percent with electron densities provided at about 9s cadence by Solar Wind Experiment instrument \cite{Ogilvie95:ssr}. We consider a period of 124 days, from October 1, 2010 to February 2, 2011, except the first 16 hours on December 7, 2010, when magnetic field measurements at $1/11$ s resolution were not available.

CSs were selected using the Partial Variance Increment (PVI) method \cite{Greco18}. We computed PVI index, ${\rm PVI}=\left(\sum_{\alpha}\Delta B_{\alpha}^2(t,\tau)/\sigma_{\alpha}^2\right)^{1/2}$, where  $\Delta B_{\alpha}(t,\tau)=B_{\alpha}(t+\tau)-B_{\alpha}(t)$ are magnetic field increments of three magnetic field components ($\alpha=X,Y,Z$) and $\sigma_{\alpha}$ are standard deviations of $\Delta {B}_{\alpha}(t,\tau)$ computed over 2h intervals, that is over a few outer correlation scales of the solar wind turbulence \cite{Matthaeus05:prl}. Coherent structures at various temporal scales correspond to non-Gaussian fluctuations with, for example, PVI$>$5. We used only PVI index computed at $\tau=1/11$ s to focus on the thinnest resolvable coherent structures. \blue{This methodology is essentially equivalent to that of \cite{Podesta17:jgr}, who used current density estimates $J\propto |\Delta {\bf B}(t,\tau)|/\tau$ with $\tau=1/11$ s to identify the most intense currents in the solar wind. The analysis of synthetic magnetic field signals with spectra typical of the solar wind showed that $J\propto |\Delta {\bf B}(t,\tau)|/\tau$ with $\tau=1/11$ s provides reasonable current density estimates \cite{Podesta17:jgr}. Note that the instrument noise does not affect the current density estimates and selection of coherent structures in our study, because the magnetometer noise level at frequencies of 0--10 Hz is less than 0.006 nT \cite{Lepping95:ssr}, which is at least six times smaller than the standard deviation, $\sigma=\left(\sum_{\alpha} \sigma_{\alpha}^{2}\right)^{1/2}$, characterizing magnetic field increments $\Delta {\bf B}(t,\tau)$ at $\tau=1/11$ s in our interval.}

There are different types of coherent structures among the non-Gaussian fluctuations \cite{Perrone16:apj}. To select CSs, each of about $2\cdot 10^{5}$ continuous clusters of points with PVI$>$5 was considered \blue{over
intervals of $\pm 1$s, $\pm 2$s, $\pm 3$s and $\pm 4$s around its center and unit vector $\textbf{\emph{x}}'$ along the direction of the magnetic field component with the largest variation} was computed for each interval using the Maximum Variance Analysis method \cite{Sonnerup&Scheible98:issi}. We visually inspected all profiles of ${\bf B}\cdot \textbf{\emph{x}}'$ and selected clusters of points with ${\bf B}\cdot \textbf{\emph{x}}'$ reversing the sign within at least one of the intervals. We then manually adjusted the boundaries\blue{, that is the regions to the left and to the right of ${\bf B}\cdot \textbf{\emph{x}}'$ reversal,} so that each boundary has a duration of at least half a second, and excluded events with substantial relative variations of the magnetic field at the boundaries. We also excluded about 10\% of events with bifurcated magnetic field profiles \cite{Gosling&Szabo08}, to focus on CSs with relatively smooth magnetic field rotation.

The final dataset includes 17,043 CSs, which is the most extensive dataset of proton kinetic-scale CSs at 1 AU. \blue{Since the CSs were selected using only PVI index with $\tau=1/11$ s, the distributions of CS parameters presented in the next section are biased toward the thinnest resolvable CSs, but this bias does not affect the key conclusions of this study (Section \ref{sec:theory}).} Each CS will be considered in local coordinate system $\textbf{\emph{xyz}}$ most suitable for describing local CS structure \cite{Knetter04,Gosling&Phan13,Phan20:apjs}: unit vector $\textbf{\emph{z}}$ is along the CS normal determined by the cross-product of magnetic fields at the CS boundaries; unit vector $\textbf{\emph{x}}$ is along $\textbf{\emph{x}}'-\textbf{\emph{z}}\cdot (\textbf{\emph{x}}'\cdot \textbf{\emph{z}})$; unit vector $\textbf{\emph{y}}$ completes the right-handed coordinate system, $\textbf{\emph{y}}=\textbf{\emph{z}}\times \textbf{\emph{x}}$. \blue{Note that vectors $\textbf{\emph{x}}$ and $\textbf{\emph{y}}$ determine the local CS surface, while vector $\textbf{\emph{z}}$ is directed across the CS surface.}

\begin{figure}[ht!]
    \centering
    \includegraphics[width=0.5\textwidth]{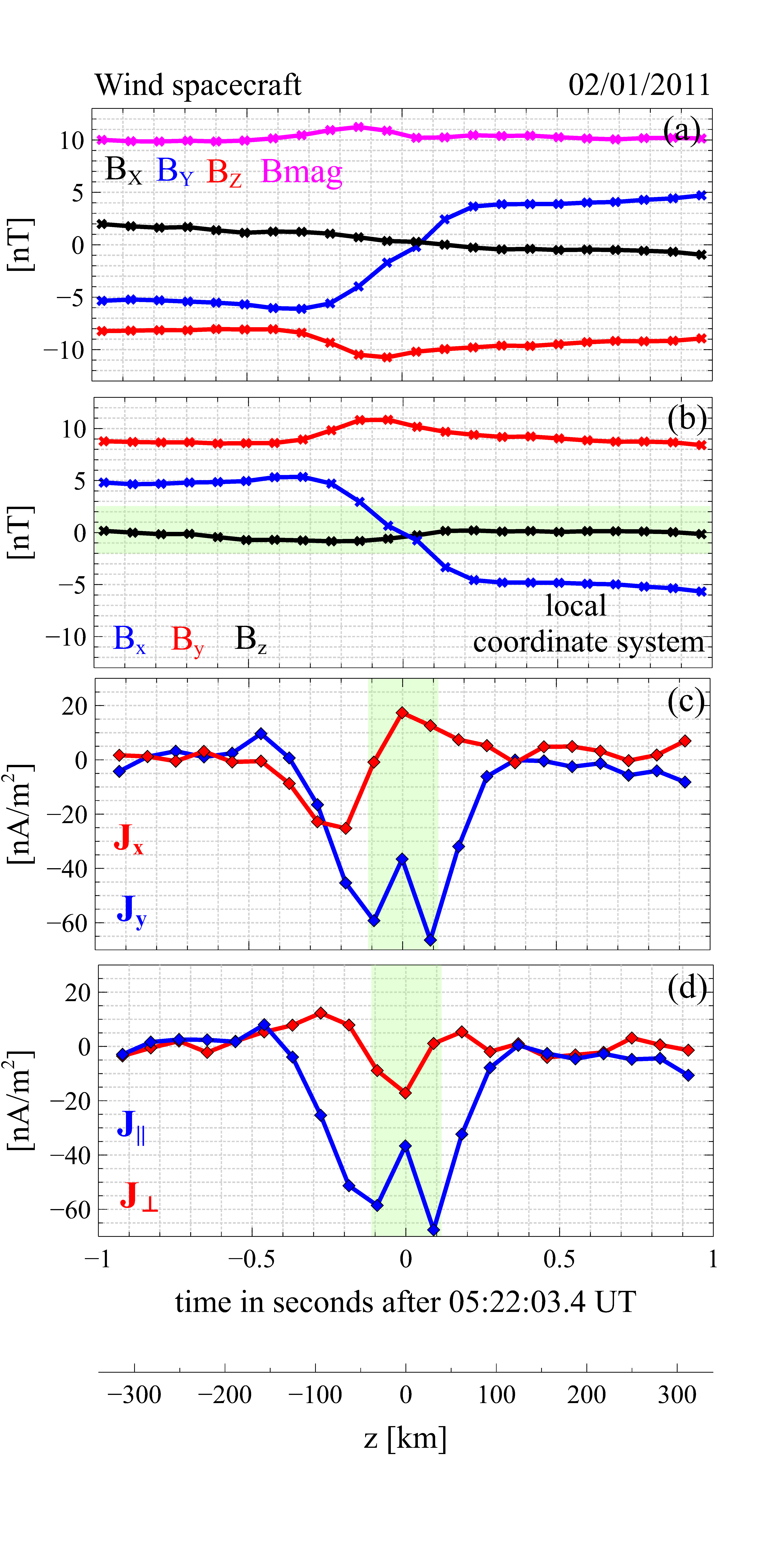}
    \caption{An example of a current sheet observed aboard Wind spacecraft on February 1, 2011 around 05:22 UT: (a) magnetic field magnitude and three magnetic field components measured at 1/11 s resolution (the Geocentric Solar Ecliptic coordinates); (b) three magnetic field components in the local coordinate system $\textbf{\emph{xyz}}$ defined in Section \ref{sec:2}; (c) current densities $J_{x}$ and $J_{y}$ computed using Eqs.(\ref{eq:J}); (d) current densities parallel and perpendicular to local magnetic field computed as $J_{||}=(J_{x}B_{x}+J_{y}B_{y})/B$ and $J_{\perp}=(J_{y}B_{x}-J_{x}B_{y})/B$. The central region of the CS, $|\;B_{x}-\langle B_{x}\rangle\;|<0.2\;\Delta B_{x}$, is indicated in panels (b)--(d), where $\langle B_{x}\rangle$ is \blue{the mean of} $B_{x}$ values at the CS boundaries, and $\Delta B_{x}$ is the absolute value of their difference. \blue{The bottom horizontal axis presents the spatial coordinate across the CS, $z=-V_{n} t$, where $t=0$ corresponds to $B_{x}=\langle B_{x}\rangle$ and $V_{n}$ is the normal component of proton flow velocity measured at the moment closest to the CS.}}
    \label{fig1}
\end{figure}

Figure \ref{fig1} presents a CS observed on February 1, 2011. Panel (a) presents the magnetic field magnitude and three components in the Geocentric Solar Ecliptic coordinate system. The magnetic field rotates across the CS through shear angle $\Delta \theta\approx 60^{\circ}$ \blue{(the angle between magnetic fields at the CS boundaries)}, while the magnetic field magnitude remains more or less constant. We characterize the magnetic field magnitude by $\langle B\rangle$ that is the \blue{mean value of magnetic field magnitudes} at the CS boundaries, and $\Delta B_{max}$ that is the difference of maximum and minimum values of the magnetic field magnitude within the CS. For the considered CS we have $\langle B\rangle \approx10$ nT and $\Delta B_{max} \approx 1.4$ nT. Panel (b) presents three magnetic field components in local CS coordinate system $\textbf{\emph{xyz}}$. The $B_{x}$ component varies from about $5$ to $-5$ nT across the CS, $B_{y}$ has similar values at the CS boundaries and a bit larger value around the $B_{x}$ reversal. The normal component $B_{z}$ is around zero at the CS boundaries in accordance with definition of the CS normal and remains small within the CS. We characterize the CS asymmetry by $\langle B_{x}\rangle$ that is \blue{the mean of $B_{x}$ values} at the CS boundaries, and the CS amplitude by the absolute value of their difference denoted as $\Delta B_{x}$. The considered CS is rather symmetric with $\langle B_{x}\rangle\approx -0.04$ nT and $\Delta B_{x}\approx 10.5$ nT. The CS central region highlighted in panel (b) corresponds to $|B_{x}-\langle B_{x}\rangle |<0.2\;\Delta B_{x}$

Panel (c) presents current densities $J_{x}$ and $J_{y}$ estimated as follows
\begin{eqnarray}
J_{y}=-\blue{\frac{1}{\mu_{0} V_{n}}}\frac{dB_{x}}{dt},\;\;\; J_{x}=\blue{\frac{1}{\mu_{0} V_{n}}}\frac{dB_{y}}{dt}
\label{eq:J}
\end{eqnarray}
where $\mu_0$ is the vacuum permeability and $V_{n}$ is the normal component of proton flow velocity at the moment closest to the CS. In this estimate we took into account that solar wind CSs are locally planar structures \cite{Knetter04,Artemyev19:grl}, \blue{so that only the normal component of the proton flow velocity matters}, and the Taylor frozen-in hypothesis is valid at spatial scales larger than electron kinetic scales \cite{Chasapis17:taylor}. The spatial coordinate along the \blue{normal, $z=-V_{n} t$ with $t=0$ corresponding to $B_{x}=\langle B_{x}\rangle$,} is shown in Figure \ref{fig1}. We estimate the CS thickness as follows
\begin{eqnarray}
\lambda=\blue{\frac{\Delta B_{x}}{2\mu_{0}\langle J_y\rangle}},
\label{eq:thickness}
\end{eqnarray}
where $\langle J_y \rangle$ is the absolute value of the current density $J_{y}$ averaged over the CS central \blue{region, $|B_{x}-\langle B_{x}\rangle |<0.2\;\Delta B_{x}$, highlighted in Figure \ref{fig1}}. For the considered CS we have $\lambda\approx 80$ km or about 1.5$\lambda_{p}$, where $\lambda_{p}$ is proton inertial length. \blue{We refer to $\lambda$ as CS thickness, but underline that strictly-speaking $\lambda$ is a half-thickness, because according to Eq. (\ref{eq:thickness}) the magnetic field and current density profiles can be approximated as $B_{x}\approx \langle B_{x}\rangle +0.5\Delta B_{x}\tanh(z/\lambda)$ and $J_{y}\approx \langle J_{y}\rangle {\rm sech}^{2}(z/\lambda)$.}


Panel (d) presents current densities parallel and perpendicular to local magnetic field computed as $J_{||}=(J_{x}B_{x}+J_{y}B_{y})/B$ and  $J_{\perp}=(J_{y}B_{x}-J_{x}B_{y})/B$. Since the perpendicular current density is statistically much smaller than the parallel current density (Section \ref{sec:theory}), we quantify the CS intensity by $J_{0}$ that is the absolute value of parallel current density $J_{||}$ averaged over the CS central region and $J_{peak}$ that is the absolute peak value of parallel current density $J_{||}$ within the CS. For the considered CS we have $J_0\approx 52$ nA/m$^{2}$ and $J_{peak}\approx 68$ nA/m$^2$. We compare the CS intensities to local Alfv\'{e}n current density $J_{A}$ and defined as the current density corresponding to the drift between protons and electrons of local Alf\'{e}n speed $V_{A}$ 
\begin{eqnarray}
J_{A}=eN_{p}V_{A},\;\;\;V_{A}=\frac{\langle B\rangle}{\left(\blue{\mu_0} N_{p}m_{p}\right)^{1/2}}
\end{eqnarray}
where $e$ and $m_{p}$ are proton charge and mass, $N_{p}$ is the closest measurement of proton density. For the considered CS, proton density was around 21 cm$^{-3}$ and we obtain $J_0\approx 0.3\;J_{A}$ and $J_{peak}\approx 0.4\;J_{A}$. In the next section we present results of the statistical analysis performed using the described methodology.




\section{Statistical Results}
 
Figure \ref{fig2} presents statistical distributions of various CS parameters. Panel (a) shows that the magnetic field magnitude across the CSs is almost constant, $\Delta B_{max}$ is less than $0.15\langle B\rangle$ for more than 95\% of the CSs, that is in accordance with the previous studies of larger-scale CSs \cite{burlaga77,Vasquez07,Artemyev19:grl}. The CSs are typically asymmetric, that is the values of $B_{x}$ at the CS boundaries are different, only about 50\% of the CSs satisfy $\langle B_{x}\rangle\lesssim 0.1\;\Delta B_{x}$. Panel (b) shows that the current densities $J_0$ are within 15 nA/m$^{2}$ for more than 95\% of the CSs. The most probable value of the current densities $J_{peak}$ is around 5 nA/m$^{2}$ that is in accordance with current density estimates in the solar wind by \cite{Podesta17:jgr}. Panel (c) shows that the thickness of the CSs is in the range between about 20 and 1000 km with the most probable value around 100 km. \blue{Note that resolving the CSs with thickness of a few tens of kilometers was possible due to relatively small normal component $V_{n}$ of the proton flow velocity for those CSs.}

Figure \ref{fig3} shows that the CSs are proton kinetic-scale structures with some fraction on sub-proton scales. Panel (a) demonstrates that the CS thickness $\lambda$ is statistically larger in solar wind plasma with larger proton inertial length $\lambda_{p}$. Similar trend is observed between the CS thickness and thermal proton gyroradius $\rho_{p}=\lambda_{p}\beta_{p}^{1/2}$, because proton beta $\beta_{p}$ is in the range between 0.4 and 2 for more than $80\%$ of the CSs in our dataset as well as in the solar wind in general \cite{Wilson18:apjs}. Panel (b) presents a statistical distribution of $\lambda/\lambda_{p}$ and shows that the thickness of the CSs is in the range between about 0.1 and 10$\lambda_{p}$ with the most probable value around $\lambda_{p}$. Similar distribution of $\lambda/\rho_{p}$ shows that the most probable value of the CS thickness is around $\rho_{p}$. \blue{Thus, the typical thickness of the CSs is from a few to ten times} smaller than of those in the previous statistical studies \cite{burlaga77,Vasquez07,Artemyev19:grl}. \blue{The resolution of these kinetic-scale CSs became possible due to higher temporal resolution of magnetic field measurements aboard Wind spacecraft.}

Figure \ref{fig4} presents analysis of the CS current density and its dependence on the CS thickness. Panel (a) shows the scatter plot of $J_0$ versus $\lambda$. We sorted the CSs into bins corresponding to different spatial scales and \blue{computed the median current density value within each bin. Panel (a) shows the median current density profile along with error bars corresponding to 15th and 85th percentiles of the current density within each bin}. The number of the CSs within each bin is shown at the bottom of panel (a). \blue{The median profile shows that the} CSs with smaller thickness tend to be more intense. The least squares fitting of all the scattered data by a power-law function reveals the following best fit
\begin{eqnarray}
J_{0}=6\; {\rm nA\; m^{-2}}\cdot \left(\frac{\lambda}{100\; {\rm km}}\right)^{-0.56},
\label{eq:1}
\label{eq:Jh}
\end{eqnarray}
\blue{which is shown in panel (a). The fitting of the 15th, 50th and 85th percentile profiles by power law functions reveals the slopes between $-0.44$ and $-0.6$, indicating thereby that the uncertainty of the best fit slope in Eq. (\ref{eq:1}) is less than 20\%.} Panel (b) presents the comparison between the peak current densities $J_{peak}$ and local Alfv\'{e}n current densities $J_{A}$. First, the CSs tend to be more intense in solar wind plasma with larger Alfv\'{e}n current density. Second, the peak current densities are statistically below local Alfv\'{e}n current density, $J_{peak}<J_{A}$ for more than 99\% of the CSs and $J_{peak}<J_{A}/2$ for 97\% of the CSs. 

Panel (c) shows there is a positive correlation between local Alfv\'{e}n speed $V_{A}$ and $J_{peak}/eN_{p}$ that is the peak value of the drift velocity between ions and electrons. Thus, the positive correlation between $J_{peak}$ and $J_{A}$ in panel (b) is not a trivial effect of plasma density variation. The physics behind the correlation of the two seemingly unrelated quantities in panel (c) will be discussed in the next section. The trends and correlations similar to those in panels (b) and (c) are also observed for the averaged current densities $J_{0}$. It is noteworthy there is no any correlation between $J_{peak}/eN_{p}$ or $J_0/eN_{p}$ and local ion-acoustic speed $(T_{e}/m_{p})^{1/2}$, where $T_{e}$ is local electron temperature. The ion-acoustic speed in our dataset is in the range from 40 to 60 km/s for more than 95\% of the CSs, while the drift velocities $J_{peak}/eN_{p}$ vary over almost two orders of magnitude.

Figure \ref{fig5} reveals remarkable scale-dependencies of normalized intensity, normalized amplitude and shear angle of the CSs. To demonstrate the scale dependence of a specific quantity, we sorted the CSs into bins corresponding to different spatial scales, and computed \blue{the median as well as 15th and 85th percentile values} of the quantity within each bin. The number of the CSs within each bin is shown at the bottom panels. Panel (a) presents the scatter plot of $J_0/J_{A}$ versus $\lambda/\lambda_{p}$. Similarly to the trend given by Eq.(\ref{eq:Jh}) in physical units, \blue{the median profile shows} that the CSs with smaller normalized thickness have larger normalized current densities. The least squares fitting of all the scattered data by a power-law function reveals the following best fit
\begin{eqnarray}
J_{0}/J_{A}=0.17\;\cdot \left(\lambda/\lambda_{p}\right)^{-0.51},
\label{eq:2}
\end{eqnarray}
\blue{which is shown in panel (a) and rather well describes the median profile. The fitting of the 15th, 50th and 85th percentile profiles by power law functions reveals the slopes in the range from $-0.45$ to $-0.55$, so that the uncertainty of the best fit slope in Eq. (\ref{eq:2}) is within about 10\%.} Panel (b) presents the scatter plot of $\Delta B_{x}/\langle B\rangle$ versus $\lambda/\lambda_{p}$. \blue{The median profile reveals} a clear scale-dependence, the CSs with larger normalized thickness have larger normalized amplitudes. The least squares fitting of all the scattered data by a power-law function reveals the following best fit
\begin{eqnarray}
\Delta B_{x}/\langle B\rangle=0.33\;\cdot \left(\lambda/\lambda_{p}\right)^{0.49},
\label{eq:3}
\end{eqnarray}
\blue{which again well describes the median profile. Note that this scaling relation could be foreseen based on Eq. (\ref{eq:2}), because $J_0\approx \Delta B_{x}/2\mu_0\lambda$ and $J_{A}=eN_{p}V_{A}=\langle B\rangle /\mu_0\lambda_{p}$. The fitting of the 15th, 50th and 85th percentile profiles reveals the slopes in the range from $0.4$ to $0.5$, so that the uncertainty of the best fit slope in Eq. (\ref{eq:3}) is within 20\%.} Since in a CS with more or less constant magnetic field magnitude, $\Delta B_{x}/\langle B\rangle$ is unambiguously related to the shear angle, we expect to observe a positive correlation and similar trend between $\Delta \theta$ and $\lambda/\lambda_{p}$. Panel (c) confirms the scale-dependence of the shear angle with the following best fit of the scattered data
\begin{eqnarray}
\Delta \theta=0.33\;\cdot \left(\lambda/\lambda_{p}\right)^{0.5}\approx 19^{\circ}\;\cdot\left(\lambda/\lambda_{p}\right)^{0.5},
\label{eq:4}
\end{eqnarray}
\blue{which well describes the median profile. The uncertainty of the best fit slope in Eq. (\ref{eq:4}) is again within 20\%.} There is a scale-dependent upper threshold on the shear angles, $\Delta \theta\lesssim 2\;\lambda/\lambda_{p}$ for more than 99\% of the CSs and $\Delta \theta\lesssim \lambda/\lambda_{p}$ for about 97\% of the CSs.

\section{Theoretical interpretation and discussion \label{sec:theory}}

\blue{We presented the analysis of proton kinetic-scale CSs in the solar wind based on the most extensive dataset collected at 1 AU. Certainly, this dataset does not include all CSs present in the considered interval. We selected the CSs using the PVI index with the minimum time increment $\tau=1/11$ s and, therefore, our dataset and statistical distributions in Figures \ref{fig2} and \ref{fig3} are biased toward the thinnest resolvable CSs. The use of PVI indexes at larger increments would be essentially equivalent to the use of magnetic field measurements at a lower resolution. Therefore, the previous studies based on a lower resolution magnetic field may demonstrate the results of using larger increments in our procedure. The studies based on magnetic field measurements at a few second resolution resolved CSs with typical thickness around 1000 km \cite{burlaga77,lepping86,soding01,Artemyev19:grl}, while the use of magnetic field measurements at 1/3 s resolution allowed resolving CSs with typical thickness around a few hundred kilometers and these CSs turned out to be one order of magnitude more abundant than CSs resolved at a few second resolution \cite{Vasquez07}. We used magnetic field measurements at 1/11 s resolution and resolved CSs with typical thickness around 100 km and averaged occurrence rate ($17,043/124\approx 140$ CSs/day) comparable with that reported by \cite{Vasquez07}. Thus, the higher resolution of magnetic field measurements allowed us to collect truly kinetic-scale CSs, which are thinner than those reported in the previous studies. Although our dataset does not include CSs of all possible spatial scales, which collection would require repeating our selection procedure with different increments $\tau$, this dataset allows us to address the structure and origin of kinetic-scale CSs in the solar wind.}

\subsection{Local structure of solar wind current sheets}

Since solar wind CSs are locally planar structures \cite{Knetter04,Artemyev19:grl}, we can describe the local magnetic field of the CSs as follows
\begin{eqnarray}
{\bf B}=B(z)\sin\theta(z)\;\textbf{\emph{x}}+B(z)\cos\theta(z)\;\textbf{\emph{y}}+B_{z}\;\textbf{\emph{z}},
\label{eq:Bmodel}
\end{eqnarray}
where $\theta(z)$ and $B(z)$ determine respectively magnetic field rotation and magnetic variation within CS, the normal component $B_{z}$ is much smaller than $B(z)$. The shear angle $\Delta \theta$ estimated in our experimental analysis corresponds to the difference of $\theta$ values at the CS boundaries. \blue{Note that Eq. (\ref{eq:Bmodel}) provides the most general description of a CS with non-zero $B_{y}$. The specific models, such as Harris CS with a constant $B_{y}$ or a force-free Harris CS, often used in reconnection simulations and theoretical studies \cite{Landi15:apjl} are special cases of Eq. (\ref{eq:Bmodel}) corresponding to particular profiles of $B(z)$ and $\theta(z)$.}

\begin{figure*}[ht!]
    \centering
    \includegraphics[width=1.0\textwidth]{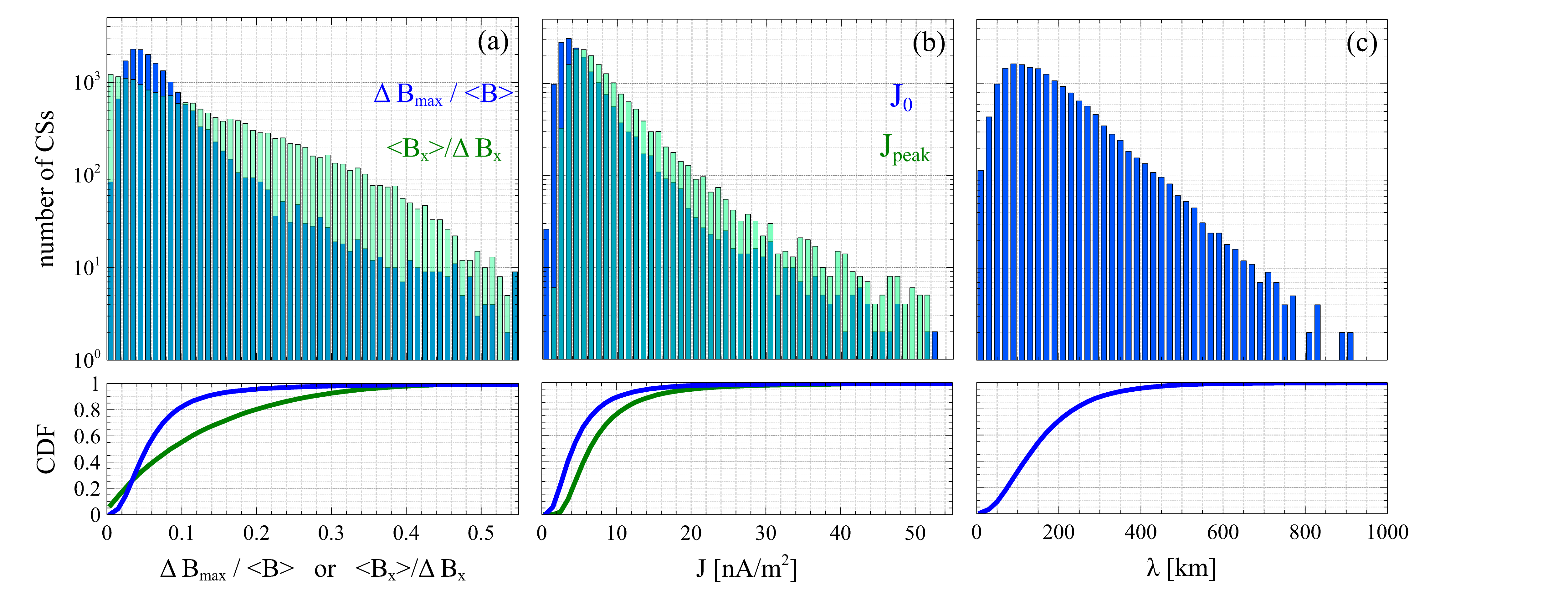}
    \caption{Statistical distributions of various parameters of 17,043 solar wind CSs: (a) $\Delta B_{max}/\langle B \rangle$, \blue{the maximum relative variation of the magnetic field magnitude within CS}, and parameter $\langle B_{x}\rangle/\Delta B_{x}$ characterizing the CS asymmetry, where $\langle B_{x}\rangle$ is the \blue{mean} of $B_{x}$ values at the CS boundaries, and $\Delta B_{x}$ is the absolute value of their difference; (b) absolute value $J_0$ of parallel current density $J_{||}$ averaged over the CS central region \blue{(see, e.g., Figure \ref{fig1})} and peak value $J_{peak}$ of parallel current density $J_{||}$ within CS; (c) CS thickness $\lambda$. The bottom panels present cumulative distribution functions (CDF) of the statistical distributions in the upper panels.}
    \label{fig2}
\end{figure*}

\begin{figure*}[ht!]
    \centering
    \includegraphics[width=1.0\textwidth]{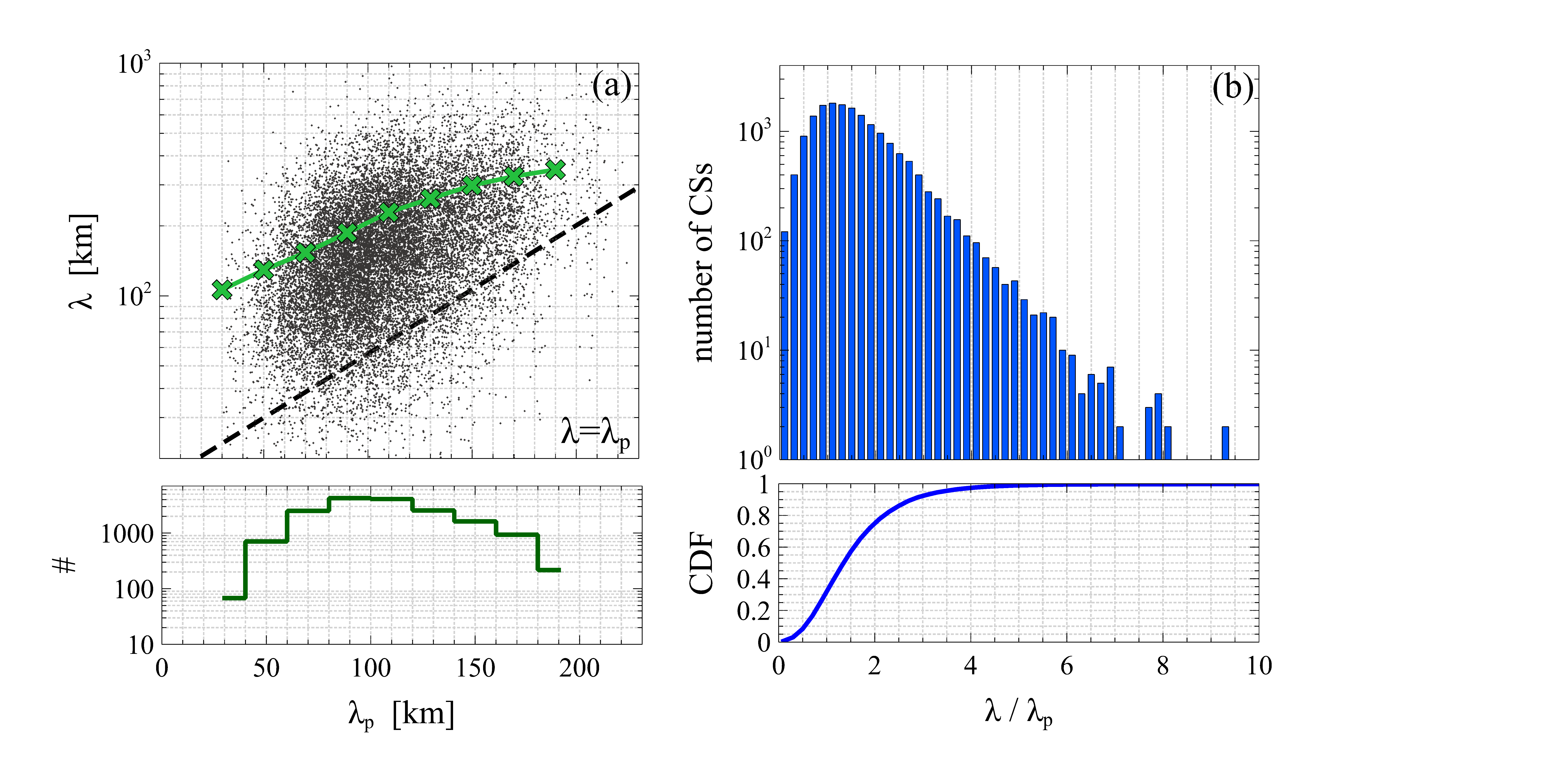}
    \caption{Panel (a) presents a scatter plot of the CS thickness $\lambda$ versus local proton inertial length $\lambda_{p}$. The green curve shows the trend that is obtained by sorting the CSs into bins corresponding to various spatial scales and computing the averaged thickness of the CSs within each bin. The number of CSs within each bin is shown in the bottom panel. \blue{The black dashed line represents $\lambda=\lambda_{p}$ for reference.} Panel (b) presents the statistical distribution of the CS thickness in units of proton inertial length. The bottom panel presents the corresponding cumulative distribution function (CDF).}
    \label{fig3}
\end{figure*}

\begin{figure*}[ht!]
    \centering
    \includegraphics[width=1.0\textwidth]{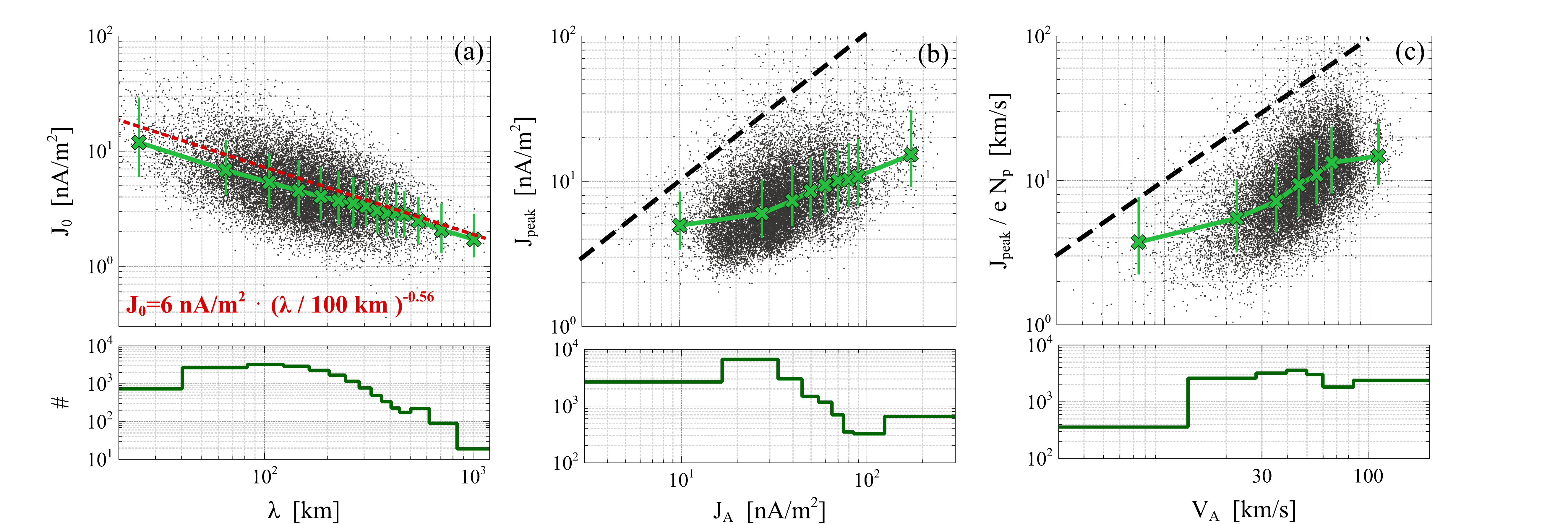}
    \caption{The scatter plots of (a) averaged current density $J_0$ versus CS thickness $\lambda$; (b) peak current density $J_{peak}$ versus local Alf\'{e}n current density $J_{A}=eN_{p}V_{A}$, where $e$ is the proton charge, $N_{p}$ is proton density, $V_{A}$ is local Alfv\'{e}n speed; (c) the peak value of the drift velocity between ions and electrons $J_{peak}/eN_{p}$ versus local Alfv\'{e}n speed $V_{A}$. For each of the scatter plots, we sorted the CSs into bins corresponding to different values of the quantity on the $x-$axis and computed \blue{median values of} the quantities on the $y-$axis for the CSs within each bin. The CSs were sorted in such a way that each bin contains a sufficiently large (more than 100) number of CSs and the number of CSs within each bin is presented in the bottom panels (only the last bin in panel (a) contains less than 100 CSs). \blue{The corresponding median profiles are presented by green curves in panels (a)--(c), while error bars correspond to 15th and 85th percentiles of the quantities within each bin. The red line in panel (a) presents the best fit of the scattered data by a power law function and the best fit parameters are indicated in the panel.} }
    \label{fig4}
\end{figure*}

\begin{figure}[ht!]
    \centering
    \includegraphics[width=0.5\textwidth]{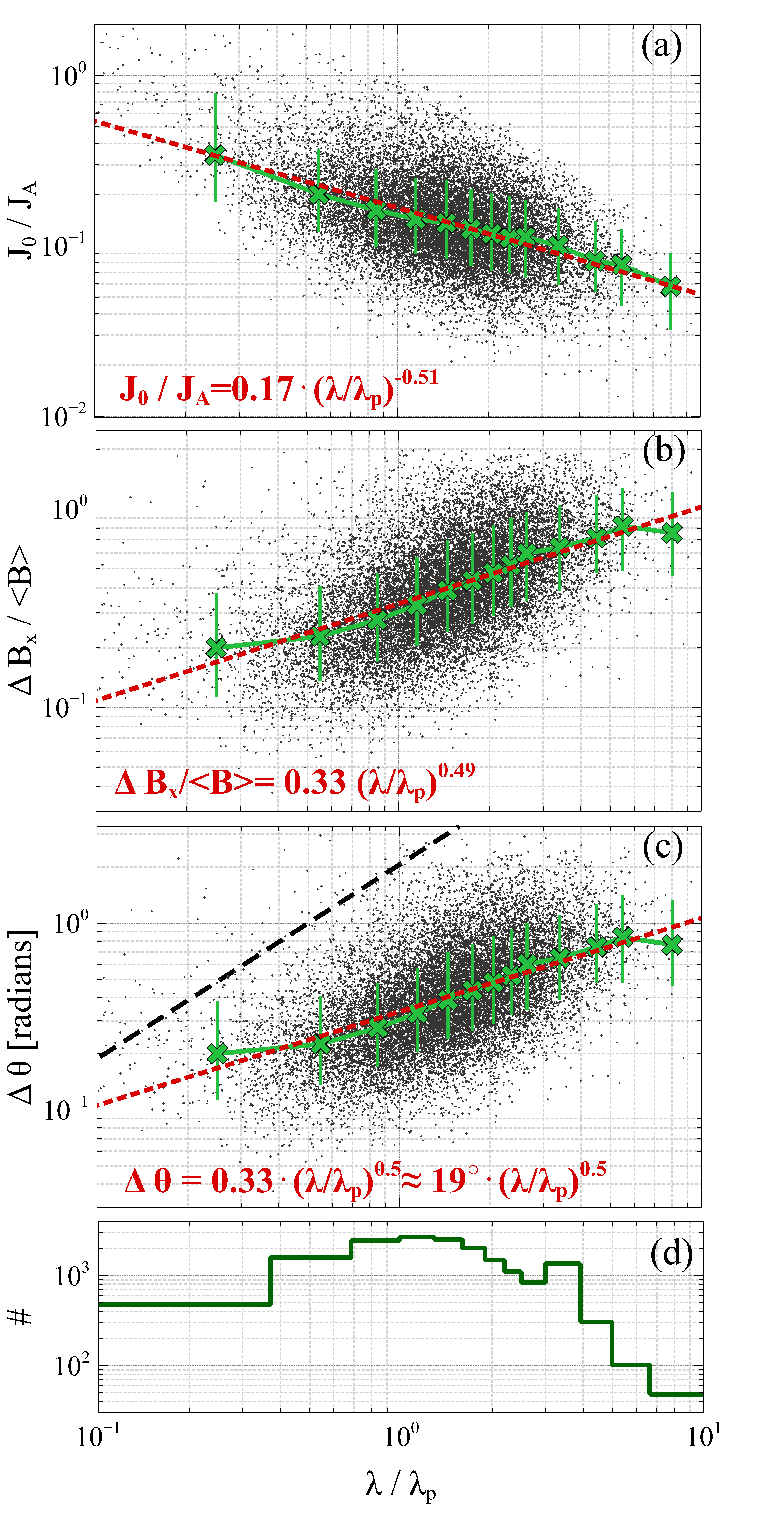}
    \caption{The scatter plots (a) the current density normalized to Alfv\'{e}n current density $J_{0}/J_{A}$ versus normalized CS thickness $\lambda/\lambda_{p}$; (b) the CS amplitude normalized to the mean magnetic field magnitude $\Delta B_{x}/\langle B\rangle$ versus $\lambda/\lambda_{p}$; (c) magnetic shear angle $\Delta \theta$ versus $\lambda/\lambda_{p}$. \blue{The CSs were sorted into bins corresponding to different values of normalized CS thickness and median values of the quantities in panels (a)--(c) were computed within each bin. The CSs were sorted in such a way that each bin contains a sufficiently large (more than 100) number of CSs within each bin and the number of CSs within each bin is presented in panel (d) (only the last bin contains less than 100 CSs). Panels (a)--(c) present the median profiles (green curves) along with error bars corresponding to 15th and 85th percentiles within each bin. The panels also indicate the best power law fits of the scattered data (red curves) along with the best fit parameters. The dashed line in panel (c) corresponds to $\Delta \theta=2\lambda/\lambda_{p}$, which corresponds to $J_{||}=J_{A}$.}}
    \label{fig5}
\end{figure}

The simple calculations show that current densities parallel and perpendicular to local magnetic field, $J_{||}=(J_{x}B_{x}+J_{y}B_{y})/\blue{B}$ and $J_{\perp}=(J_{y}B_{x}-J_{x}B_{y})/\blue{B}$, determine respectively magnetic field rotation and magnitude variation within CS
\begin{eqnarray}
J_{||}=\frac{B}{\blue{\mu_0}}\frac{d\theta}{dz},\;\;\;\;\;\;
J_{\perp}=\frac{1}{\blue{\mu_0}}\frac{dB}{dz}
\label{eq:Jmodel}
\end{eqnarray}
The amplitudes of parallel and perpendicular current densities can be then estimated as $J_{||}\approx (c/4\pi) \langle B\rangle \Delta \theta/2\lambda$ and $J_{\perp}\approx (c/4\pi) \Delta B_{max}/2\lambda$. 
\blue{First, these estimates show that} the current density limitation $J_{||}\lesssim J_{A}$ (Figures \ref{fig4}b) is equivalent to the scale-dependent upper threshold on shear angles, $\Delta \theta\lesssim 2\lambda/\lambda_{p}$ (Figure \ref{fig5}c). \blue{Second, they show that} the relatively small variations of the magnetic field magnitude within the CSs (Figure \ref{fig2}a) imply that the current density is dominated by the parallel component. The ratio $J_{\perp}/J_{||}\approx \Delta B_{max}/\langle B\rangle \Delta \theta$ is less than 0.3 for more than 95\% of the CSs (not shown). \blue{The small variations of the magnetic field magnitude also imply that the plasma pressure variation across the CSs is much smaller than the magnetic field pressure. The pressure balance $p+B^{2}/8\pi={\rm const}$ shows that the maximum variation of the plasma pressure across CS is $\Delta p_{\rm max}\approx \langle B\rangle \Delta B_{max}/4\pi$, while its ratio to the typical magnetic field pressure is $\Delta p/p_{B}\approx 2\Delta B_{max}/\langle B\rangle$. According to Figure \ref{fig2}a, this ratio does not exceed 0.3 for more than 95\% of the CSs. The dominance of the parallel current density and relatively small variations of the plasma pressure show that the kinetic-scale CSs in the solar wind are statistically more or less force-free.}



\subsection{Current sheets and turbulence}

The turbulence in the solar wind is dominated by magnetic field fluctuations highly oblique to mean magnetic field, with a power law spectrum $E_{k_{\perp}}\propto k_{\perp}^{-\nu}$, where $\nu\approx 5/3$ in the inertial range, $k_{\perp}\lambda_{p}\lesssim 1$, and \blue{$\nu\approx 2.8$} at $k_{\perp}\lambda_{p}\gtrsim 1$ \cite{Chen16:jpp}. The root-mean-square amplitude $\delta b_{\lambda}$ of turbulent fluctuations at spatial scale $\lambda$ can be then estimated as $\delta b_{\lambda}\propto (E_{k_{\perp}}\Delta k_{\perp})^{1/2}\propto \lambda^{(\nu-1)/2}$, where we took into account that $\Delta k_{\perp}\propto k_{\perp}\propto \lambda^{-1}$. The corresponding current density is $j_{\lambda}\propto \delta b_{\lambda}/\lambda\propto \lambda^{(\nu-3)/2}$. In the inertial range, $k_{\perp}\lambda_{p}\lesssim 1$, we expect
\begin{eqnarray}
\delta b_{\lambda}\propto \lambda^{1/3}, \;\;\;j_{\lambda}\propto \lambda^{-2/3}
\end{eqnarray}
while at $k_{\perp}\lambda_{p}\gtrsim 1$ we expect
\begin{eqnarray}
\delta b_{\lambda}\propto \lambda^{\blue{0.9}}, \;\;\;j_{\lambda}\propto \lambda^{-\blue{0.1}}
\end{eqnarray}
Thus, the root-mean-square amplitudes of turbulent magnetic field and current density fluctuations should be scale-dependent \cite{Schekochihin09,Boldyrev&Perez12}. Since magnetic shear angle $\delta \theta_{\lambda}$ is proportional to $\delta b_{\lambda}$, it should be similarly scale-dependent.  

We have found that amplitudes, current densities and shear angles of proton kinetic-scale CSs are scale-dependent in a fashion similar to that of turbulent fluctuations. Moreover, the CS amplitude scales with the CS thickness as $\Delta B_{x}/\langle B\rangle \propto (\lambda/\lambda)^{0.49}$, so that the power law index is between $1/3$ and \blue{$0.9$} expected for turbulent fluctuations at scales above and below proton kinetic scales. Similarly, the CS intensity scales with the CS thickness as $J_0/J_{A}\propto (\lambda/\lambda_{p})^{-0.51}$ with the power law index between $-2/3$ and \blue{$-0.1$}. These scale-dependencies are strong indications that proton kinetic-scale CSs are produced locally by plasma turbulence. \blue{This conclusion is not affected by 20\% uncertainty of the slopes characterizing the scale-dependent properties of the CSs.}

\blue{The critical question is whether the scale-dependencies revealed in Figure \ref{fig5} would be affected if we included all kinetic-scale CSs present in the considered interval, rather than used the dataset biased toward the thinnest resolvable CSs. The strong indication that the revealed scale-dependencies would not be affected is that they} are consistent with \blue{the scale-dependent properties} of magnetic field rotations in the solar wind \cite{Zhdankin12:apjl,Chen15:mnras}. These studies showed that statistical distributions of angles $\alpha(\tau)$ between magnetic fields ${\bf B}(t)$ and ${\bf B}(t+\tau)$ in the solar wind behave similarly at various temporal scales. More precisely, $\alpha(\tau)$ normalized to its mean value $\langle \alpha\rangle$ is described by a universal log-normal distribution independent of $\tau$. In turn, the mean value scales with $\tau$ as $\langle \alpha\rangle\propto \tau^{0.46}$, so that magnetic field rotation angles are larger at larger spatial scales (see also \cite{Perri12:prl}). This is qualitatively and even quantitatively consistent with the scale-dependence of magnetic shear angles across the CSs, $\Delta \theta\propto (\lambda/\lambda_{p})^{0.5}$, \blue{revealed in our analysis}. \blue{Thus, we believe that inclusion of kinetic-scale CSs, which could be selected using PVI indexes at larger increments, would not affect our conclusion that kinetic-scale CSs are produced by turbulence cascade.}

\subsection{Current density limitation mechanisms}

\blue{We have found that the peak value of the current density within CS is correlated with local Alfv\'{e}n current density $J_{A}$, while the drift velocity between ions and electrons is correlated with local Alfv\'{e}n speed $V_{A}$. In principle, the positive correlation between these quantities could be expected, because $J_{A}$ and $V_{A}$ are natural units of the current density and velocities in MHD and Hall-MHD turbulence simulations \cite{Papini19:apj}. The intriguing property though is that the peak value $J_{peak}$ of the current density does not statistically exceed $J_{A}$ and, actually, even $J_{A}/2$}. There are several scenarios capable of explaining the observed current density limitation. The first scenario is that once the parallel current density exceeds local Alfv\'{e}n current some instability may lead to current density limitation or a CS destruction. One of the known instabilities is the so-called Alf\'{e}n instability that was considered by \cite{Voitenko95:solphys} and \cite{Bellan99} for a force-free CS at low plasma betas. The relevance of this instability to solar wind CSs at realistic plasma betas requires a separate analysis. The ion-acoustic instability is highly unlikely to be relevant to current density limitation, because we have found no correlation between the ion-electron drift velocity $J_{peak}/eN_{p}$ and local ion-acoustic speed (see also \cite{Boldyrev15:apj}). The second scenario is that once the ion-electron drift velocity becomes comparable with local Alfv\'{e}n speed, more electrons can be in the Landau resonance \blue{with ambient turbulence \cite{TenBarge13}, which potentially} results in electron scattering and current density limitation.


\subsection{Our dataset and assumptions of the single-spacecraft analysis}

Wind spacecraft allowed us to select the most extensive dataset of proton kinetic-scale CSs in the solar wind not disturbed by the Earth's bow shock. In our single-spacecraft analysis CS normals were determined by the cross-product of magnetic fields at the CS boundaries, which works relatively well according to the multi-spacecraft study by \cite{Knetter04}. Multi-spacecraft studies would certainly allow more accurate estimates of CS normals, but the available multi-spacecraft missions often probe the solar wind disturbed by the Earth's bow shock and require careful data selection.

We repeated the analysis presented in this Letter for the current densities and thicknesses estimated via Eqs. (\ref{eq:J}) and (2), but using proton flow velocity magnitude in the $\textbf{\emph{yz}}$ plane rather than along CS normals $\textbf{\emph{z}}$. The so-obtained quantities represent lower estimates of the current densities and upper estimates of the CS thicknesses, which are independent of the exact knowledge of CS normals. We found similar scale-dependencies with only slightly different fitting parameters, $J_0\approx 6\;{\rm nA\;m^{-2}}\cdot (\lambda/100\;{\rm km})^{-0.49}$, $J_0/J_{A}\approx 0.14\cdot (\lambda/\lambda_{p})^{-0.4}$, $\Delta B_{x}/\langle B\rangle \approx 0.28\cdot (\lambda/\lambda_{p})^{0.6}$, and $\Delta \theta\approx 15^{\circ}\cdot (\lambda/\lambda_{p})^{0.62}$. Thus, we feel confident that the results of our single-spacecraft analysis reflect realistic properties of proton kinetic-scale CSs in the solar wind.

\section{Conclusions}

The results of this Letter can be summarized as follows
\begin{enumerate}
    \item The proton kinetic-scale CSs in the solar wind are statistically force-free and typically asymmetric.
    
    \item The current density \blue{within the CSs} is scale-dependent with CSs of smaller thickness being more intense. The magnetic field amplitude and magnetic shear angle are scale-dependent as well. 

    \item The \blue{drift velocity} between electrons and ions in the CSs tends to be larger in the solar wind with larger Alfv\'{e}n speed. The current density does not statistically exceed local Alfv\'{e}n current density.
\end{enumerate}

Based on these observations we argue that proton kinetic-scale CSs in the solar wind are produced locally by turbulence, and some mechanism\blue{, either CS instability or scattering of electrons by ambient turbulence,} should keep the current density below local Alfv\'{e}n current density.

{\bf Acknowledgments:}
The work of I.V. was supported by the Russian Science Foundation, grant No. 21-12-00416. The work of T.P. was supported by NASA Living With a Star grant \#80NSSC20K1781. The work of A.A. was supported by NASA Living With a Star grant \#80NSSC20K1788. The work of K.A. was supported by National Science Foundation grant No. 2026680. I.V. thanks Ajay Lotekar and Rachel Wang for discussions. The data used in the analysis are publicly available at https://cdaweb.gsfc.nasa.gov/pub/data/wind/.

\bibliographystyle{unsrt}
\bibliography{full}  

\end{document}